\documentclass{sf2a-conf2011}
\usepackage{graphicx}
\usepackage{hyperref}
\usepackage[]{natbib}  
\usepackage[cyr]{aeguill}
\usepackage{epstopdf}

\def\BibTeX{{\rm B\kern-.05em{\sc i\kern-.025em b}\kern-.08em
    T\kern-.1667em\lower.7ex\hbox{E}\kern-.125emX}}
\bibpunct{(}{)}{;}{a}{}{,}  

\begin{document}

\TitreGlobal{SF2A 2011}


\title{SPADES: a Stellar PArameters DEtermination Software}

\runningtitle{SPADES }

\author{H. Posbic }\address{GEPI, Observatoire de Paris, CNRS, Universit\'{e} Paris Diderot, Place Jules Janssen, 92190, Meudon, France}

\author{D. Katz$^1$}

\author{E. Caffau}\address{Zentrum f$\ddot u$r Astronomie der Universit$\ddot a$t Heidelberg, Landessternwarte, K$\ddot o$nigstuhl 12, 69117, Heidelberg, Germany}

\author{P. Bonifacio$^1$}

\author{L. Sbordone$^2$}

\author{A. Gomez$^1$}

\author{F. Arenou$^1$}



\setcounter{page}{237}

\index{Posbic, H}
\index{Katz, D}
\index{Caffau, E}
\index{Bonifacio, P}
\index{Sbordone, L}
\index{Gomez, A}
\index{Arenou, F}


\maketitle


\begin{abstract}
With the large amounts of spectroscopic data available today and the very large surveys to come (e.g. Gaia), 
the need for automatic data analysis software is unquestionable. We thus developed an automatic spectra 
analysis program for the determination of stellar parameters: radial velocity, effective temperature, 
surface gravity, micro-turbulence, metallicity and the elemental abundances of the elements present in the spectral range. Target stars for 
this software should include all types of stars. The analysis method relies on a line by line comparison of
 the spectrum of a target star to a library of synthetic spectra. The idea is built on the experience acquired
 in developing the TGMET (\cite{Katz98} and \cite{Soubiran03}) ETOILE \citep{Katz01} and Abbo \citep{Bonifacio03} 
software.The method is presented and the performances are illustrated with 
GIRAFFE-like simulated spectra with high resolution (R = 25000), with high and low signal to noise ratios 
(down to SNR= 30). These spectra should be close to what could be targeted by the Gaia-ESO Survey (GCDS).
\end{abstract}

\begin{keywords}
stellar parameters, spectra analysis, Giraffe
\end{keywords}


\section{Introduction}
 
One of the major applications of spectroscopy is the determination of stellar parameters like the radial velocity (Vr), effective temperature (Teff), 
surface gravity (log(g)) , micro-turbulence ($\xi$) , metallicity ([Fe/H]) and chemical abundances ([X/H]). 
The present and future large spectroscopic surveys are going to significantly increase the number of spectroscopic data to be analysed. 
A few examples are the Gaia-ESO Survey with about $160'000$ stars to be observed, Gaia with about $2'000'000$ stars to be analysed for chemical abundances \citep{Katz04}, RAVE with some $400'000$ stars observed so far (\cite{RAVE11} and \cite{Siebert11}) etc.
The space mission Gaia will provide the largest survey ever, and that, in the decade to come.
 To analyse these quantities of data, automatic spectra analysis software is needed.  
Different families of software exist. A few examples are software like TGMET (\cite{Katz98} and \cite{Soubiran03}, ETOILE \citep{Katz01}, MATISSE \citep{A.Recio-Blanco06}, Abbo \citep{Bonifacio03}  etc...
The work presented is about the development of a new automatic stellar spectra anaylysis software.
In its first version the software will be optimised for medium-resolution Giraffe spectra (VLT) and will thus be tested on Giraffe like spectra. 
The software is called SPADES (Stellar PArameters DEtermination Software) and is coded with Java. 

\section{SPADES}

\subsection{ General idea}
The software is based on the comparison between observed spectra and a grid of synthetic spectra (with known parameters).
Contrary to many existing software, in SPADES, the comparison between spectra is not made over all the spectrum but around pre-selected lines. 
Another particularity is that the determination of the stellar parameters does not use any equivalent width mesures but is based on profile fitting methods.
Another important characteristic of SPADES is that it determines elemental abundances. 
The general idea is as follows: \\

For each parameter to be determined, one or several methods of determination (diagnostics) are available. One is chosen by the user. The list of diagnostics for each parameter are:

\begin{itemize}
 \item Vr: cross-correlation in direct space with a template
 \item Teff: excitation equilibrium or Balmer lines profile fitting.
 \item log(g): ionization equilibrium or strong lines (e.g. MgIb green triplet) profile fitting.
 \item $[$Fe$/$H$]$: Fe lines profile fitting.
 \item $[$X$/$H$]$ : X lines profile fitting.
 \item $\xi$: empiric calibration or nulling the $\Delta W = f($reduced equivalent widths$)$ function slope. $\Delta W$ being the residuals of the difference between the observed and synthetic line.

\end{itemize}

The diagnostics so far implemented and tested are detailed in the next section.
Each of the parameters, Teff, log(g),  $[Fe/H]$ and $\xi$ atmospheric parameters is determined one by one assuming all the others known. The process is iterated until convergence.
The elemental chemical abundances are then determined.\\

\subsection{ Diagnostics}
\subsubsection {Effective temperature : Teff}

\begin{figure}[ht!]
 \centering
\includegraphics[width=0.4\textwidth,clip]{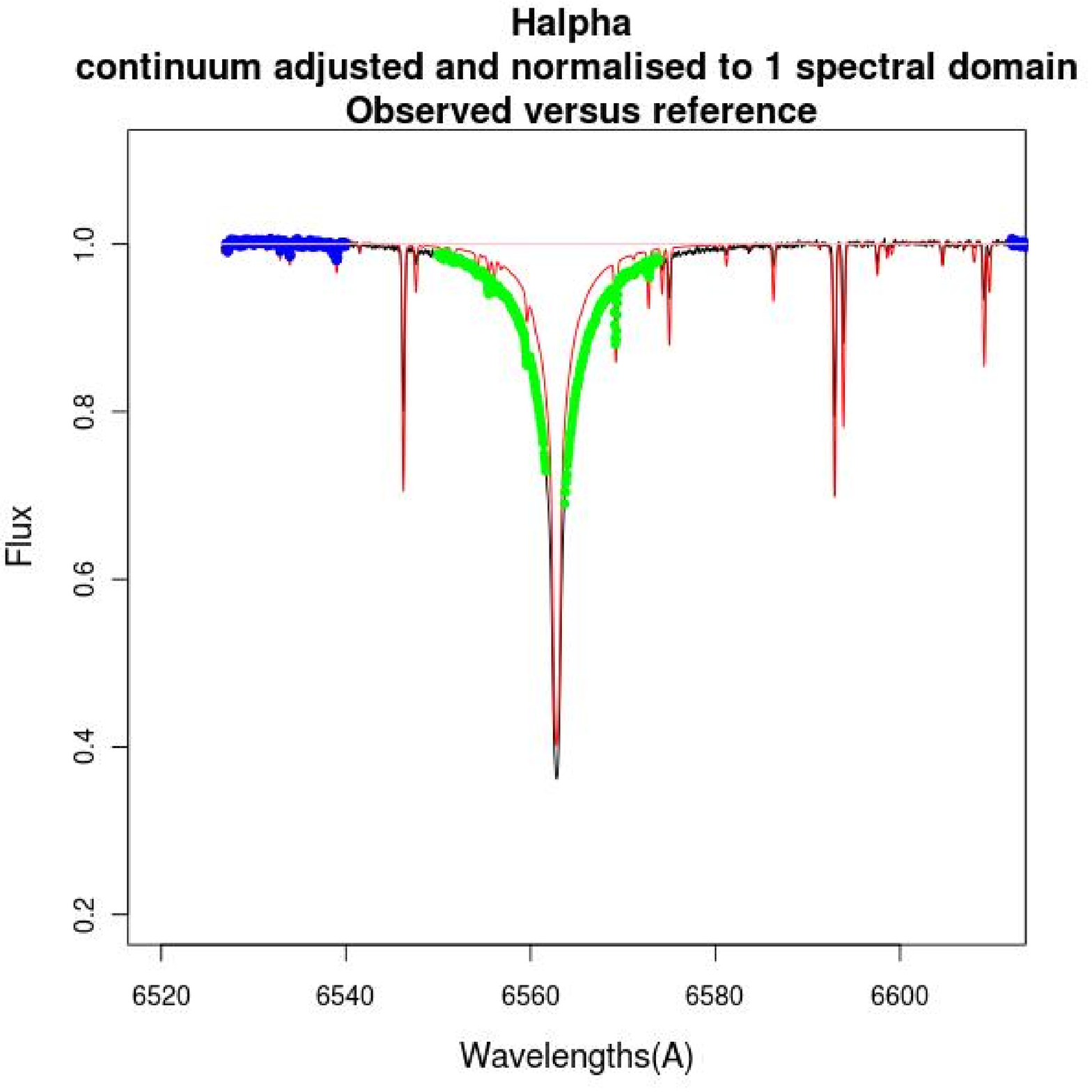}%
 \includegraphics[width=0.4\textwidth,clip]{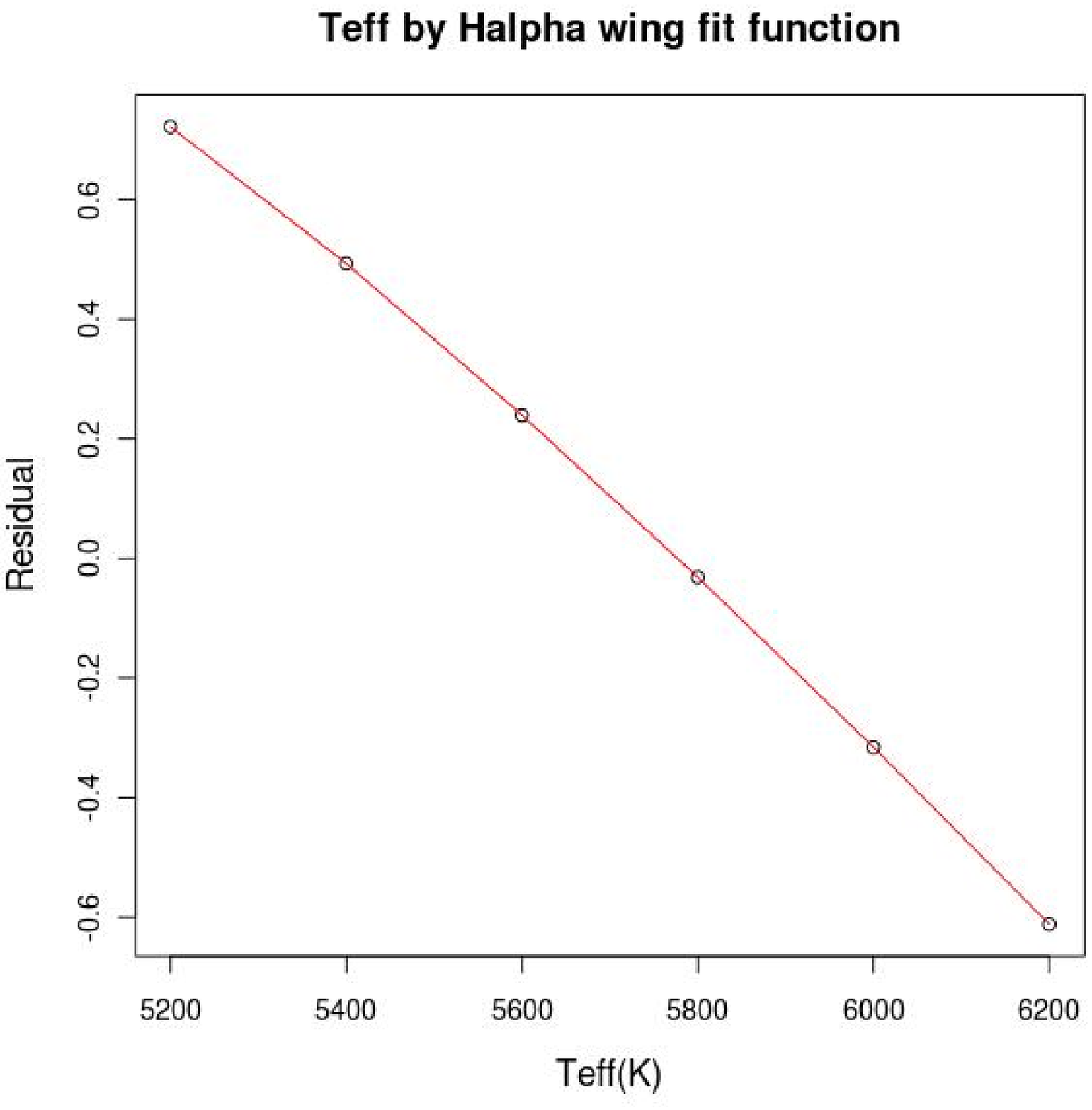}      
  \caption{{\bf Left:} The reference spectrum H$\alpha$ spectral domain in red is superposed to the studied spectrum H$\alpha$ spectral domain in black. The continuum pixels are in blue in Fig.~\ref{author1:fig1} . The H$\alpha$ wings pixels are in green.
	   {\bf Right:} The residuals as function of the reference spectra Teff. The result Teff is the value that nulls this function. }
  \label{author1:fig1}
\end{figure}

\subparagraph {H$\alpha$ wings fit method}
:\\
The first step is defining the reference grid to be used: a 1D (in the parameters space) reference spectra grid is defined. 
This grid varies over Teff only, the other parameters beeing fixed to their input values. 
This grid is read or calculated by interpolation based on a pre-calculated reference spectra grid. 
The analysis is limited to the H$\alpha$ line, more generarely to a spectral range around this line.
This range will be now called the ``H$\alpha$ spectral domain''. 
For each reference spectrum (Teff value), the H$\alpha$ spectral domain continuum is fitted to the studied H$\alpha$ spectral domain continuum.
An example of superposed continuum fitted H$\alpha$ spectral domains is presented in Fig.~\ref{author1:fig1} { \bf Left}.
The spectral domains differ by their Teff. 

The wavelengths ranges used for the continuum fit are pre-defined. The corresponding pixels are in blue. 
The green pixels represent the H$\alpha$ wings. The wavelength limits of the wings are also pre-defined. 
Over the H$\alpha$ wings pixels the residual is calculated as such: 

$s = \sum_{pixels} (x_{obs} - \tilde x_{ref})$
 
That is done for all the spectra in the 1D reference spectra grid. 
$\ s = f(Teff_{ref})$ is thus constructed. 
An example of this function is presented in Fig.~\ref{author1:fig1} { \bf Right}.

The result Teff is the one that nulls this function. 

\begin{figure}[ht!]
\centering
\includegraphics[width=0.6\textwidth,clip]{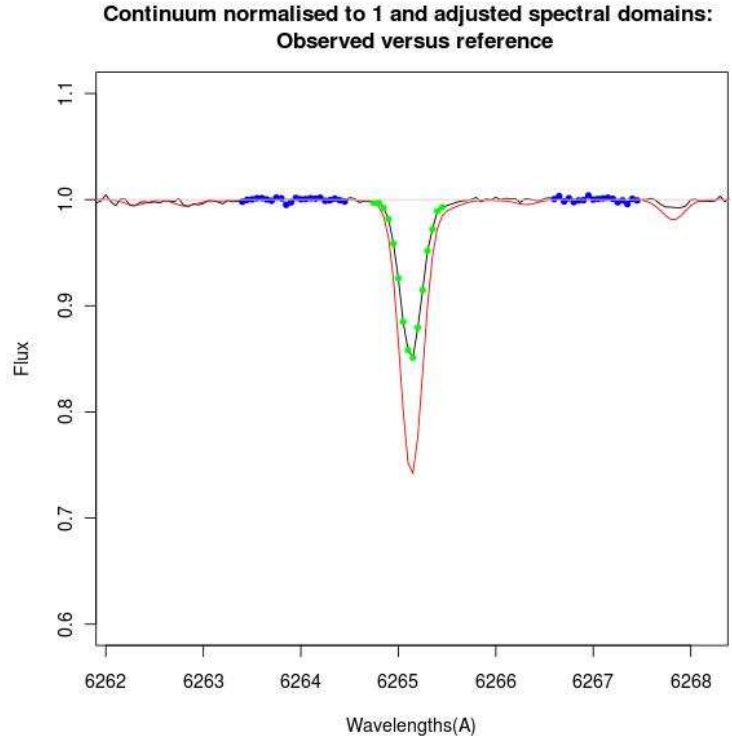}      
\caption{In red the spectral domain of a FeI line in the reference spectrum superposed to the studied spectral domain of the same line, in black. 
The studied and the reference spectra are at different Teffs which explains the difference between the line.
 The continuum pixels are in blue. The line pixels are in green.}
\label{author1:fig2}
\end{figure}

\subparagraph {Excitation equilibrium} %
  :\\ In this method a list of pre-selected FeI lines is used. 
As in the previous method, a 1D (in the parameters space) reference spectra grid is defined with Teff values varying around the input value while the other parameters are 
fixed to their input values. Around each central wavelength a spectral domain is cut in the reference and observed spectra. 
The line and continuum limits are determined automatically. 
For each reference spectrum, the spectral domains around the pre-selected lines are continuum fitted to 
the continuum of their corresponding spectral domain in the studied spectrum. 
An example of superposed, continuum fitted spectral domains is given in Fig.~\ref{author1:fig2}

For each reference spectral domain, and over the line pixels (green pixels) the residual is measured as such:
 
$\Delta W = \sum_{pixels} - (x_{obs} -\tilde x_{ref})$

\begin{figure}[ht!]
 \centering
\includegraphics[width=0.4\textwidth,clip]{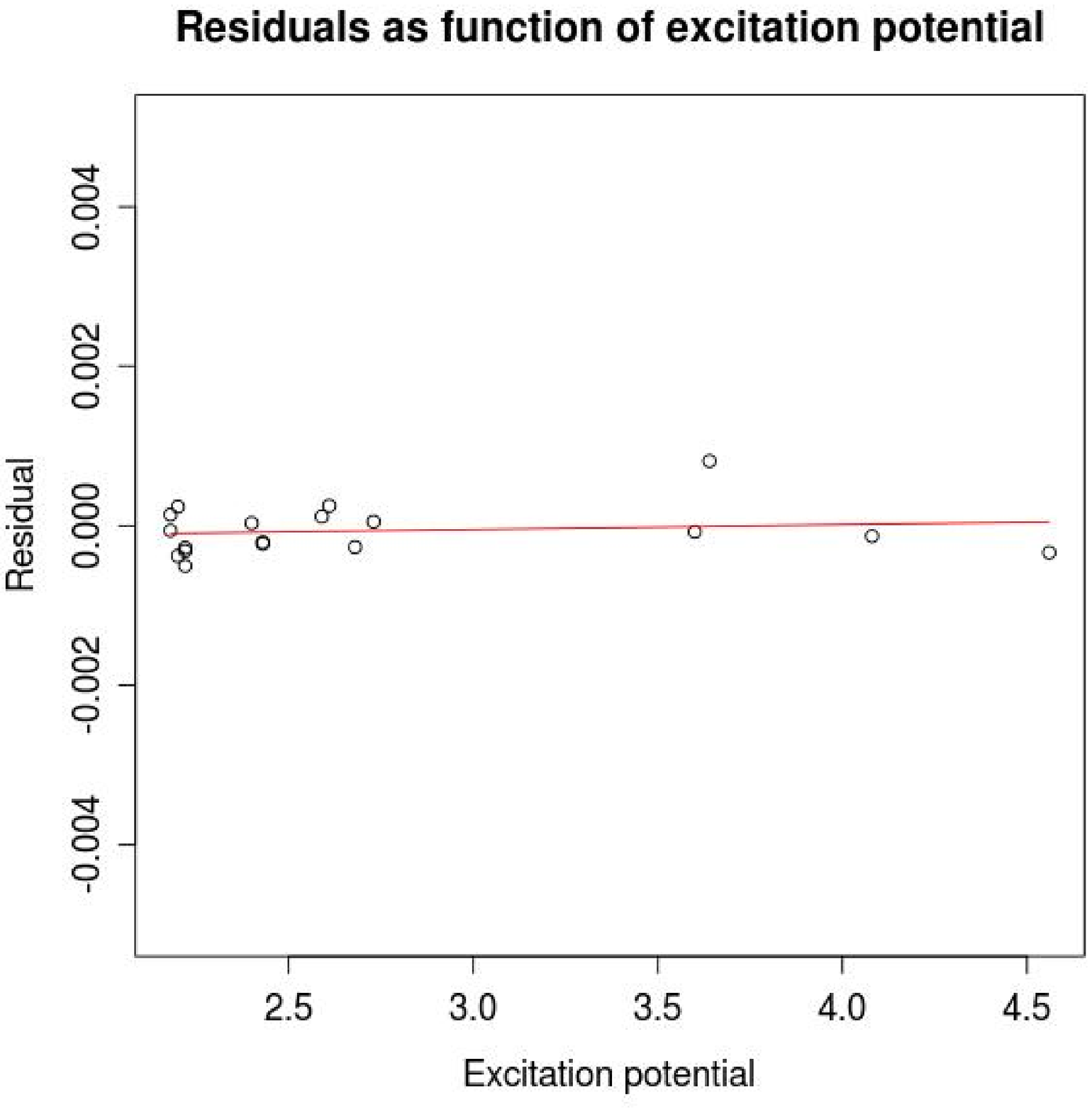}      
\includegraphics[width=0.4\textwidth,clip]{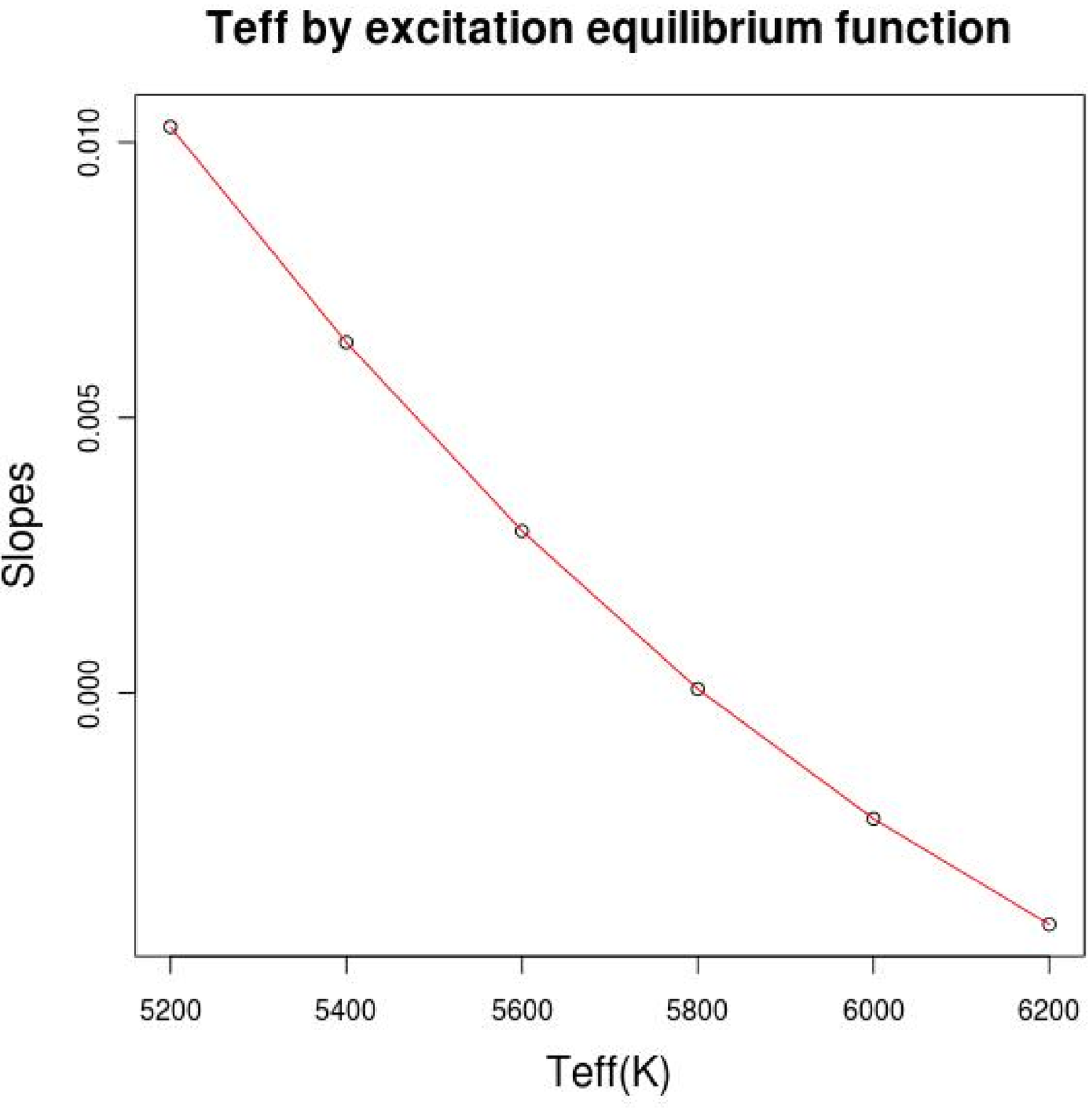}   
\caption{{\bf Left:} Line residuals ($\Delta W_{n}$) as function of excitation potentials $\xi_{n}$ for the reference spectrum with Teff = 5800 K. {\bf Right : } The slopes a such as $ \Delta W_{n} = a*\xi_{n} + b$ as a function of the reference spectra Teff. The result Teff is the value that nulls this function. }
\label{author1:fig3}
\end{figure}

This measurement is done for all the used lines. Let $\Delta W_{n}$ be the residual of the nth line. For each reference spectrum, the lines $\Delta W_{n}$ are plotted as a function of the 
lines respective excitation potentials $\xi_{n}$. 
An example of this function is given in Fig.~\ref{author1:fig3} {\bf Left}

For each reference spectrum (Teff value) the slope a of this function is measured ($ \Delta W_{n} = a*\xi_{n} + b$).
Another function is then constructed: the $a = f(Teff)$ function.
An example of this function is presented in Fig.~\ref{author1:fig3} {\bf Right}. 

The result Teff is the one that nulls the  $a = f(Teff)$ function. 

\subsubsection {Gravity: log(g) }
\subparagraph {Ionisation equilibrium}
: \\ In this method the measurement made over the FeI and the FeII lines to be used are: \\
 
$\Delta W_{FeI} = \sum_{pixels} - (x_{obs} -\tilde x_{ref})$  and  $\Delta W_{FeII} = \sum_{pixels} - (x_{obs} -\tilde x_{ref})$ \\ 
 
The diagnostic to be analysed is : 
 
$\Delta = \bar {\Delta W_{FeI}} - \bar {\Delta W_{FeII}}$ \\
 \\where $\bar {\Delta W_{FeI}}$ (respectively $\bar {\Delta W_{FeII}}$)
are the mean of the $\Delta W_{FeI_{n}}$s (respectively the  $\Delta W_{FeII_{n}}$s)for all the FeI (respectively the FeII) lines. 
The result log(g) nulls the $\Delta$ as a function of the reference spectra logg function. 

\subsubsection {Metallicity and elemental abundances: [X/H]}
\subparagraph {Profile fit}
: \\ In this method the measurement over the n lines of the X element to be used is :  $s^2 = \sum_{pixels} \frac{(x_{obs} - \tilde x_{ref})^2}{\sigma^2}$ 

The diagnostic to be analysed over all the X element lines is:  $S^2 =\sum_{lines} \sum_{pixels} \frac{(x_{obs} - \tilde x_{ref})^2}{\sigma^2}$

The result $[$X$/$H$]$ is the one that nulls the $S^2$ as function of the reference spectra $[$X$/$H$]$ function.

\subsection{ Tests and perfomances}
SPADES was tested by Monte-Carlo over synthetic spectra with resolution $ R = 25000 $, effective temperature $Teff = 5800K$, gravity $ log(g) = 4.40 $, metallicity $ [Fe/H]= -1.0$, with individual abundances of Ca and Ni : $[Ca/H] = 0.0 $ and $ [Ni/H] = 0.0$.
The tests were made for 2 signal to noise ratios (SNR): 30 and 100 (200 reference spectra for each SNR).
The values of the dispersions at $1-\sigma$ of the residuals (difference between the estimated and the real value) for each parameter are as follows: \\
\\
\begin{tabular}{|l|l|l|}
\hline
		& $SNR = 30$ 	& $SNR = 100$  \\
\hline
Teff (K)        &     31       	&   	9	\\
log(g)       	&     0.14    	&   	0.05	\\
$[$Fe$/$H$]$ (dex)  &     0.04	&	0.0013	\\
$[$Ca$/$H$]$ (dex)  &     0.03	&	0.009	\\
$[$Ni$/$H$]$ (dex)  &     0.05	&	0.017	\\
 \hline
\end{tabular}
\\
\\
\\
The mean results of these Monte-Carlo runs for each parameter show no bias. 
The dispersions are acceptable. Actually, for the Teff determination using the H$\alpha$ method, the dispersion is at SNR of 100 (respectively 30) about 10 times (respectively 3 times) smaller than the systematic error (estimated by \cite{Cayrel11}) linked to the physics behind the models used. 
We note that, of course, the H$\alpha$ line is not always available for use: one reason is that it can simply not be in the spectral domain used, another is that this Teff determination method cannot be used for all stars (cool stars for example).
The excitation equilibrium method is then used.

\section{Future work}
The future work to be done on the software is :
\\- On the fly reference grid calculation: dynamic call of the SYNTHE software for calculating the reference grid directly from SPADES (completed at the time of writing the proceedings) 
\\- Fix a method for determining micro-turbulence
\\- Determine the external errors (as opposed to the internal errors determined by Monte-Carlo). One of the methods will be the test on known stars (e.g Sun)
\\- In its first version, the software will be fine-tuned to analyze medium to high resolution GIRAFFE spectra of Thick Disk stars .

\bibliographystyle{aa}  
\bibliography{posbic} 

\end{document}